\documentclass[mathleft]{an}
\usepackage{graphicx}
\usepackage{times}
\usepackage{natbib}
\bibpunct{(}{)}{;}{a}{}{,}
\def\beq{\begin{eqnarray}}
\def\eeq{\end{eqnarray}}

\providecommand{\eprint}[2]{\url{#2}}

\begin{document}

\title{The XMM Cluster Outskirts Project (X-COP)}

\author{D. Eckert\inst{1}\fnmsep\thanks{Corresponding author:
  \email{Dominique.Eckert@unige.ch}} \and  S. Ettori\inst{2,3} \and E. Pointecouteau\inst{4,5} \and S. Molendi\inst{6} \and S. Paltani\inst{1} \and C. Tchernin\inst{7} \and The X-COP collaboration
}
\titlerunning{Instructions for authors}
\authorrunning{D. Eckert et al.}
\institute{
Department of Astronomy, University of Geneva, ch. d'Ecogia 16, 1290 Versoix, Switzerland
\and 
INAF - Osservatorio Astronomico di Bologna, Via Ranzani 1, 40127 Bologna, Italy
\and
INFN, Sezione di Bologna, viale Berti Pichat 6/2, 40127 Bologna, Italy
\and
CNRS; IRAP; 9 Av. colonel Roche, BP 44346, F-31028 Toulouse cedex 4, France
\and
Universit\'e de Toulouse; UPS-OMP; IRAP; Toulouse, France
\and
INAF - IASF-Milano, Via E. Bassini 15, 20133 Milano, Italy
\and
Center for Astronomy, Institute for Theoretical Astrophysics, Heidelberg University, Philosophenweg 12, 69120 Heidelberg, Germany
}

\received{}
\accepted{}
\publonline{later}


\abstract{Galaxy clusters are thought to grow hierarchically through the continuous merging and accretion of smaller structures across cosmic time. In the Local Universe, these phenomena are still active in the outer regions of massive clusters ($R>R_{500}$), where the matter distribution is expected to become clumpy and asymmetric because of the presence of accreting structures. We present the \emph{XMM-Newton} Cluster Outskirts Project (X-COP), which targets the outer regions of a sample of 13 massive clusters ($M_{500}>3\times10^{14}M_\odot$) in the redshift range 0.04-0.1 at uniform depth. The sample was selected based on the signal-to-noise ratio in the \emph{Planck} Sunyaev-Zeldovich (SZ) survey with the aim of combining high-quality X-ray and SZ constraints throughout the entire cluster volume. Our observing strategy allows us to reach a sensitivity of $3\times10^{-16}$ ergs cm$^{-2}$ s$^{-1}$ arcmin$^{-2}$ in the [0.5-2.0] keV range thanks to a good control of systematic uncertainties. The combination of depth and field of view achieved in X-COP will allow us to pursue the following main goals: \emph{i)} measure the distribution of entropy and thermal energy to an unprecedented level of precision; \emph{ii)} assess the presence of non-thermal pressure support in cluster outskirts; \emph{iii)} study the occurrence and mass distribution of infalling gas clumps. We illustrate the capabilities of the program with a pilot study on the cluster Abell 2142.}

\maketitle

\section{Introduction}
In the hierarchical structure formation paradigm, galaxy clusters are expected to form through the continuous merging and accretion of smaller structures \citep[see][for a review]{kravtsov12}. In the local Universe, such processes should be observable in the outer regions of massive clusters, where galaxies and galaxy groups are infalling for the first time and smooth material is continuously accreted from the surrounding cosmic web. 

The hot plasma in galaxy clusters is expected to be heated to high temperatures ($10^7-10^8$ K) through shocks and adiabatic compression at the boundary between the free-falling gas and the virialized intra-cluster medium \citep[ICM,][]{tozzi00}. The thermodynamical properties of the gas retain information on the processes leading to the thermalization of the gas in the cluster's potential well, which is encoded in the gas entropy $K=kTn_e^{-2/3}$. Gravitational collapse models predict that the entropy of stratified cluster atmospheres increases steadily with radius, following a power law with index $\sim1.1$ \citep{voit05,borgani05,sembolini15}. However, non-gravitational processes induce an additional injection of entropy and can therefore be traced through the departures from the theoretical predictions \citep{chaudhuri12}. Such departures have been observed for a long time in cluster cores, where gas cooling and feedback from supernovae and active galactic nuclei are important \citep[e.g.][]{david96,ponman99,pratt10}. More recently, several works reported a deficit of entropy in massive clusters around the virial radius \citep[see][for a review]{reiprich13}, which has been interpreted as a lack of thermalization of the ICM induced, e.g., by an incomplete virialization of the gas \citep[e.g.][]{kawa,bonamente13,ichikawa13}, non-equilibration between electrons and ions \citep{hoshino}, non-equilibrium ionization \citep[e.g.][]{fujita08}, or weakening of the accretion shocks \citep{lapi}. However, these models have received little support from cosmological simulations so far \citep[e.g.][]{va10kp,lau15,avestruz15}. 

The gas content of infalling dark-matter halos interacts with the ICM and is stripped from its parent halo through the influence of the ram pressure applied by the ICM of the main cluster. This process is expected to be the main mechanism through which the infalling gas is heated up and virialized into the main dark-matter halo \citep{gunn72,vollmer01,heinz03,roediger15} and it is believed to be key to the evolution of the cluster galaxy population by quenching rapidly the star formation activity in clusters \citep{roediger08,bahe15}. Recent observational evidence suggest that thermal conduction in the ICM is strongly inhibited \citep[e.g.][]{gaspari13,sanders14}. The long conduction timescale therefore delays the virialization of the stripped, low-entropy gas inside the potential well of the main cluster \citep{eckert14b,degrandi16}, which causes the ICM in the outer regions of massive clusters to be clumpy \citep{mathiesen,nagai,vazza12c}. Since the X-ray emissivity depends on the squared gas density, inhomogeneities in the gas distribution lead to an overestimation of the mean gas density \citep{nagai,simionescu,eckert15}, which biases the measured entropy low. This effect needs to be taken into account when measuring the entropy associated with the bulk of the ICM. In addition, large-scale accretion patterns in the direction of the filaments of the cosmic web induce asymmetries in the gas distribution \citep[e.g.][]{vazza11a,e12,roncarelli13}. Such filaments are expected to host the densest and hottest phase of the warm-hot intergalactic medium \citep[e.g.][]{cen99,dave01,eckert15b}, which are expected to account for most of the missing baryons in the local Universe.

In this paper, we present the \emph{XMM-Newton} cluster outskirts project (X-COP), a very large programme on \emph{XMM-Newton} that aims at advancing significantly our knowledge of the physical conditions in the outer regions of galaxy clusters ($R>R_{500}$\footnote{For a given overdensity $\Delta$, $R_\Delta$ is the radius for which $M_\Delta/(4/3\pi R_\Delta^3)=\Delta\rho_c$}). X-COP targets a sample of 13 massive, nearby clusters selected on the basis of their high signal-to-noise ratio (SNR) in the \emph{Planck} all-sky survey of Sunyaev-Zeldovich \citep[SZ,][]{sz} sources \citep{planckpsz1,planckpsz2}. In the recent years, the progress achieved in the sensitivity of SZ instruments allowed to extend the measurements of the pressure profile of galaxy clusters out to the virial radius and beyond \citep{planck5,sayers13}. The high SNR in the \emph{Planck} survey ensures a detection of the SZ effect from our targets well beyond $R_{500}$. X-COP provides a uniform 25 ks mapping of these clusters out to $R_{200}$ and beyond, with the aim of combining high-quality X-ray and SZ imaging throughout the entire volume of these systems. 

\section{Sample selection}

To implement the strategy presented above, we selected a list of the most suitable targets to conduct our study. The criteria used for the selection are the following:
\begin{enumerate}
\item \textbf{SNR $\mathbf{>12}$ in the PSZ1 catalog \citep{planckpsz1}:} This condition is necessary to target the most significant \emph{Planck} detections and ensure that the SZ effect from all clusters be detected beyond $R_{500}$;
\item \textbf{Apparent size $\mathbf{\theta_{500}>10}$ arcmin:} Given the limited angular resolution of our reconstructed SZ maps ($\sim7$ arcmin), this condition ensures that all the clusters are well-resolved, such that the contamination of SZ flux from the core is low beyond $R_{500}$ ;
\item \textbf{Redshift in the range $\mathbf{0.04<z<0.1}$:} This criterion allows us to cover most of the azimuth out to $R_{200}$ with 5 \emph{XMM-Newton} pointings (one central and four offset) whilst remaining resolved by \emph{Planck};
\item \textbf{Galactic $\mathbf{N_H<10^{21}}$ cm$\mathbf{^{-2}}$:} Since we are aiming at maximizing the sensitivity in the soft band, this condition makes sure that the soft X-ray signal is weakly absorbed.
\end{enumerate}

This selection yields a set of the 15 most suitable targets for our goals. We excluded three clusters (A2256, A754, and A3667) because of very complicated morphologies induced by violent merging events, which might hamper the analysis of the \emph{Planck} data given the broad \emph{Planck} beam. The remaining 12 clusters selected for our study are listed in Table \ref{tab:master}, together with their main properties. A uniform 25 ks mapping with \emph{XMM-Newton} was performed for 10 of these systems in the framework of the X-COP very large programme (Proposal ID 074441, PI: Eckert), which was approved during \emph{XMM-Newton} AO-13 for a total observing time of 1.2 Ms. The remaining 2 systems (A3266 and A2142) were mapped by \emph{XMM-Newton} previously. Although the available observations of A3266 do not extend all the way out to $R_{200}$, they are still sufficient for some of our objectives and we include them in the present sample. Finally, we add Hydra A/A780 to the final sample. While the SZ signal from this less massive cluster is not strong enough to be detected beyond $R_{500}$, a deep, uniform \emph{XMM-Newton} mapping exists for this system \citep[see][for more details]{degrandi16}.

Our final sample therefore comprises 13 clusters in the mass range $2\times10^{14}<M_{500}<10^{15}M_\odot$ and X-ray temperature $3<kT<10$ keV. In Table \ref{tab:master} we also provide the values of the central entropy $K_0$ from the ACCEPT catalog \citep{cavagnolo}, which is an excellent indicator of a cluster's dynamical state \citep{hudson10}. According to this indicator, five of our clusters are classified as relaxed, cool-core systems ($K_0<30$ keV cm$^2$), while the remaining eight systems are dynamically active, non-cool-core clusters.

\begin{table*}
\caption{\label{tab:master}Master table presenting the basic properties of the X-COP sample.}
\begin{center}
{
\begin{tabular}{lcccccccccc}
\hline
Name & $z$ & SNR & $L_{X,500}$ & $kT_{\rm vir}$ & $Y_{500}$ & $M_{500}$ & $R_{500}$ & $\theta_{500}$  & $K_{0}$ & Ref\\
\, &  & \emph{Planck}  &  [$10^{44}$ergs s$^{-1}$] & [keV] & [$10^{-3}$ arcmin$^2$] & [$10^{14} M_\odot$] & [kpc] & [arcmin] & [keV cm$^2$]\\
\hline
\hline
A2319 & 0.0557 & 49.0 & $5.66\pm0.02$ & $9.60_{-0.30}^{+0.30}$ & 43.17 & 10.56 & 1525 & 23.49 & $270.23 \pm4.83$ & 2 \\
A3266$^{\star}$ & 0.0589 & 40.0 & $3.35\pm0.01$ & $9.45_{-0.36}^{+0.35}$ & 23.52 & 10.30 & 1510 & 22.09 & $72.45 \pm 49.71 $ & 1 \\
A2142$^\star$ & 0.090 & 28.4 & $8.09\pm0.02$ & $8.40_{-0.76}^{+1.01}$ & 18.54 & 8.51 & 1403 & 13.92 & $68.06 \pm 2.48$ & 1 \\
A2255 & 0.0809 & 26.5 & $2.08\pm0.02$ & $5.81_{-0.20}^{+0.19}$ & 11.17 & 4.94 & 1172 & 12.80 & $529.10 \pm 28.19$ & 1 \\
A2029 & 0.0766 & 23.2 & $6.94\pm0.02$ & $8.26_{-0.09}^{+0.09}$ & 12.66 & 8.36 & 1399 & 16.08 & $10.50 \pm 0.67$ & 1 \\
A85 & 0.0555 & 22.8 & $3.74\pm0.01$ & $6.00_{-0.11}^{+0.11}$ & 16.97 & 5.24 & 1205 & 18.64 & $12.50 \pm 0.53 $ & 1 \\
A3158 & 0.059 & 19.8 & $2.01\pm0.01$ & $4.99_{-0.07}^{+0.07}$ & 10.62 & 3.98 & 1097 & 16.03 & $166.01 \pm 11.74$ & 1 \\
A1795 & 0.0622 & 19.3 & $4.43\pm0.01$ & $6.08_{-0.07}^{+0.07}$ &  6.43 & 5.33 & 1209 & 16.82 & $18.99 \pm 1.05$ & 1 \\
A644 & 0.0704 & 17.3  & $3.40\pm0.01$ & $7.70_{-0.10}^{+0.10}$ & 7.22 & 7.55 & 1356 & 16.82 & $132.36 \pm 9.15$ & 3 \\
A1644 & 0.0473 & 16.1 & $1.39\pm0.01$ & $5.09_{-0.09}^{+0.09}$ & 13.96 & 4.12 & 1115 & 20.02 & $19.03 \pm 1.16$ & 1 \\
RXC J1825 & 0.065 & 15.2 & $1.38\pm0.01$ & $5.13_{-0.04}^{+0.04}$ & 8.39 & 4.13 & 1109 & 14.81 & $217.93\pm6.33$ & 4 \\
ZwCl 1215 & 0.0766 & 12.8$^\dagger$ & $2.11\pm0.01$ & $6.27_{-0.29}^{+0.32}$ & - & 5.54 & 1220 & 14.01 & $163.23 \pm 35.62$ & 1 \\
A780$^\star$ & 0.0538 & -$^{\ddagger}$ & $2.25\pm0.01$ & $3.45_{-0.09}^{+0.08}$ & - & 2.75 & 872 & 13.87 & $13.31 \pm 0.66$ & 1 \\
\hline
\end{tabular}
}
\end{center}
\textbf{Column description:} 1. Cluster name. The clusters identified with an asterisk were mapped prior to X-COP. Abbreviated names: RXC J1825.3+3026, ZwCl 1215.1+0400, A780/Hydra A ; 2. Redshift (from NED); 3. Signal-to-noise ratio (SNR) in the \emph{Planck} PSZ2 catalog \citep{planckpsz2}. $^\dagger$In PSZ1 \citep{planckpsz1}, but not in PSZ2 as it falls into the PSZ2 point source mask (see Table E.4 in \citet{planckpsz2}). The SNR expected in PSZ2 from Eq. 6 and Table 3 in \citet{planckpsz2} is about 16. $^{\ddagger}$Below both PSZ1 and PSZ2 detection threshold; 4. Luminosity in the [0.5-2] keV band (rest frame); 5. Virial temperature; 6. Integrated $Y$ parameter from the PSZ2 catalog; 7. Mass within an overdensity of 500, estimated using the $M-T$ relation of \citet{arnaud05}; 8. Corresponding value of $R_{500}$ (in kpc); 9. Apparent size of $R_{500}$ in arcmin; 10. Central entropy $K_0$, from \citet{cavagnolo}; 11. Reference for the cluster temperature. 1: \citet{hudson10}; 2: \citet{molendi99}; 3: \citet{cavagnolo} ; 4: This work (in prep.)
\end{table*}

\section{Abell 2142: a pilot study}

\begin{figure}
\resizebox{\hsize}{!}{\includegraphics{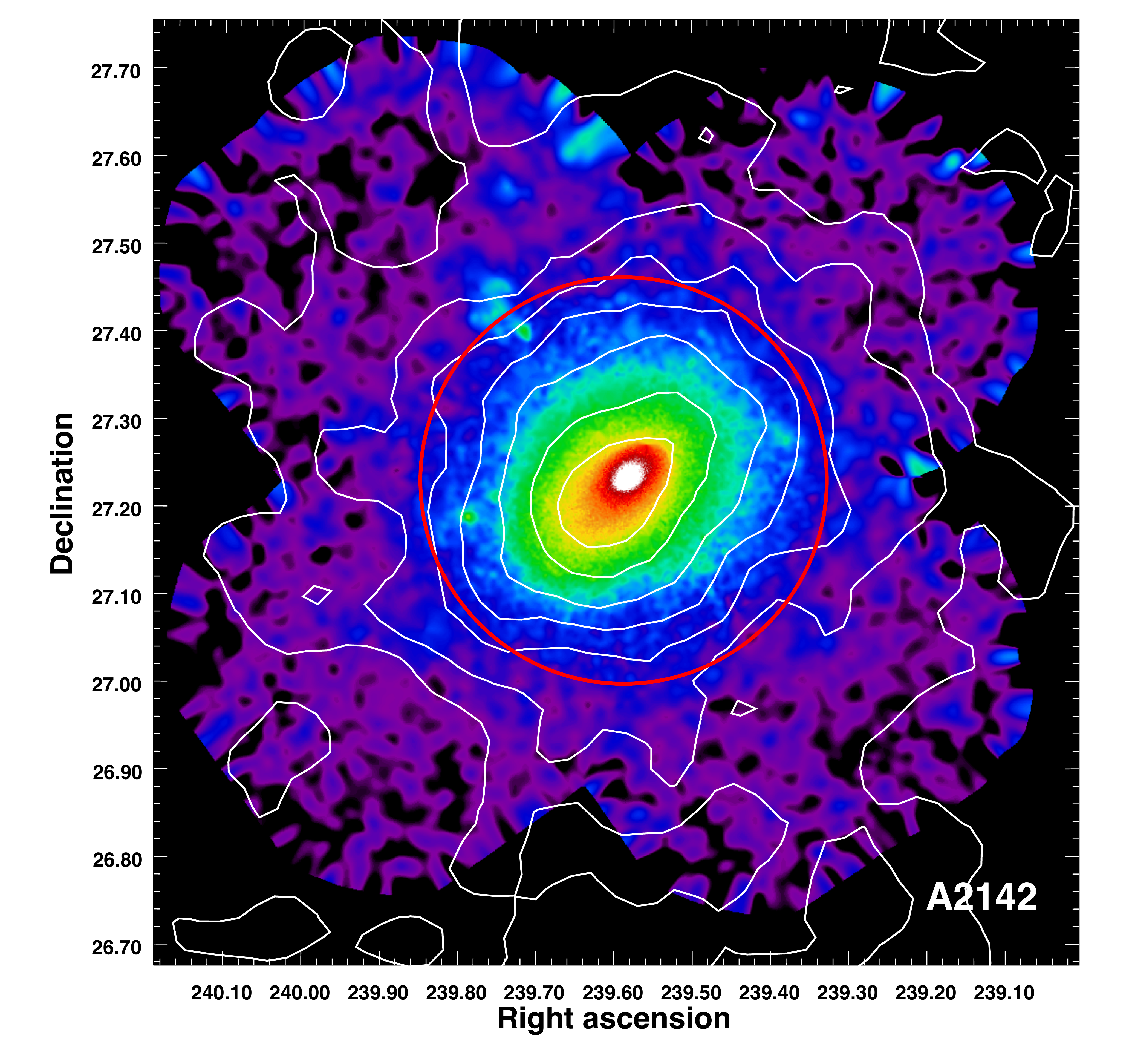}}
\caption{Adaptively smoothed, background subtracted \emph{XMM-Newton} image of Abell 2142 in the [0.7-1.2] keV band. The corresponding \emph{Planck} Compton-parameter contours are shown in white. The contour levels correspond to 1, 3, 5, 7, 10, 15, 20, 30, 40, and 50$\sigma$. The red circle indicates the estimated value of $R_{500}\sim1,400$ kpc.}
\label{fig:xmm2142}
\end{figure}

Abell 2142 \citep[$z=0.09$,][]{owers11} is the first cluster for which the X-COP observing strategy was applied. In \citet{tchernin16} we presented our analysis of this system out to the virial radius, highlighting the capabilities of X-COP. The results of this program are summarized here. In Fig. \ref{fig:xmm2142} we show an adaptively smoothed, background subtracted \emph{XMM-Newton}  mosaic image of Abell 2142 in the [0.7-1.2] keV range, with \emph{Planck} contours overlayed.

\subsection{\emph{XMM-Newton} surface-brightness profile} 

We developed a new technique to model the \emph{XMM-Newton} background by calculating two-dimensional models for all the relevant background components: the non X-ray background (NXB), the quiescent soft protons (QSP), and the cosmic components. To validate our background subtraction technique, we analyzed a set of 21 blank fields totaling 1.3 Ms of data. The analysis of this large dataset yields a flat surface-brightness profile, with a scatter of 5\% around the mean value. This analysis allows us to conclude that the background level can be recovered with a precision of 5\% in the [0.7-1.2] keV band \citep[see Appendix A and B of][]{tchernin16}.

To measure the average surface-brightness profile free of the clumping effect, we applied the technique developed in \citet{eckert15}. Namely, in each annulus we computed the distribution of surface-brightness values by applying a Voronoi tessellation technique \citep{cappellari03} and estimated the median surface brightness from the resulting distribution. The median of the surface brightness distribution was found to be a robust estimator of the mean gas density \citep{zhuravleva13}, unlike the mean of the distribution, which is biased high by the presence of accreting clumps. The ratio between the mean and the median can thus be used as an estimator of the clumping factor,

\begin{equation}C=\frac{\langle\rho^2\rangle}{\langle\rho\rangle^2},\end{equation}

\noindent where $\langle\cdot\rangle$ denotes the mean over radial shells \citep[see][for a validation of this technique using numerical simulations]{eckert15}. This technique allows us to excise all clumps down to the size of the Voronoi bins, which in the case of Abell 2142 corresponds to 20 kpc.

\begin{figure}
\resizebox{\hsize}{!}{\includegraphics[width=0.5\textwidth]{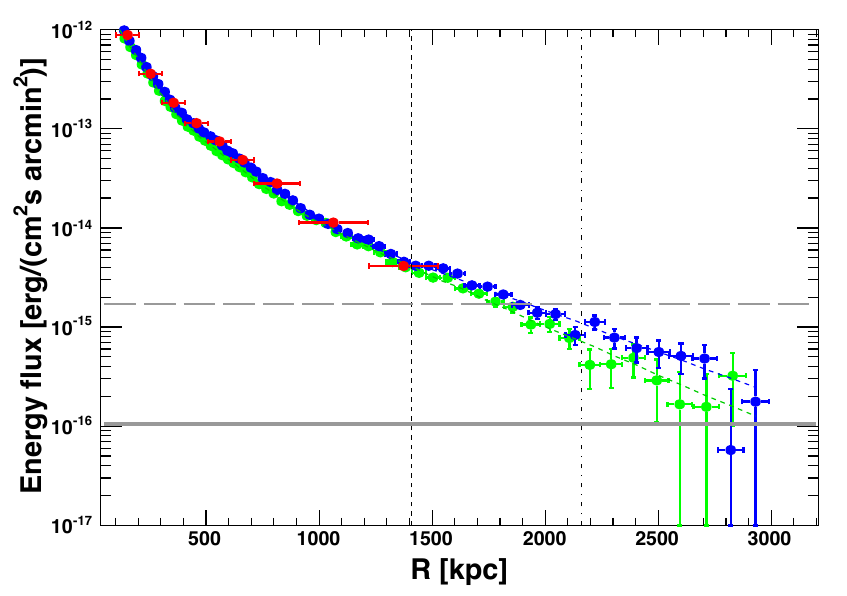}}
\caption{Mean (blue) and median (green) \emph{XMM-Newton} surface-brightness profiles of Abell 2142 in the [0.7-1.2] keV band. The red data points show the results obtained using a spectral analysis. The horizontal dashed and solid lines show the sky background level and the level of systematics, respectively.}
\label{fig:sb}
\end{figure}

In Fig. \ref{fig:sb} \citep[reproduced from][]{tchernin16} we show the mean and median surface-brightness profiles of Abell 2142. The median profile is clearly below the mean at large radii, which highlights the importance of clumping in cluster outskirts. A significant X-ray signal is measured out to 3 Mpc from the cluster core ($\sim2R_{500}$), beyond which the systematics dominate. Note the significant improvement over previous \emph{XMM-Newton} studies, which were typically limited to the region inside $R_{500}$ \citep[e.g.][]{lm08,pratt07}.

\subsection{\emph{Planck} SZ pressure profile}

Abell 2142 is one of the strongest detections in the \emph{Planck} survey, with an overall signal-to-noise ratio of 28.4 using the data from the full \emph{Planck} mission (see Table \ref{tab:master}. A significant SZ signal is observed as well out to 3 Mpc from the cluster core, and it can be readily transformed into a high-quality pressure profile. For the details of the \emph{Planck} analysis procedure we refer to \citet{planck5}. 

\begin{figure}
\resizebox{\hsize}{!}{\includegraphics[width=0.5\textwidth]{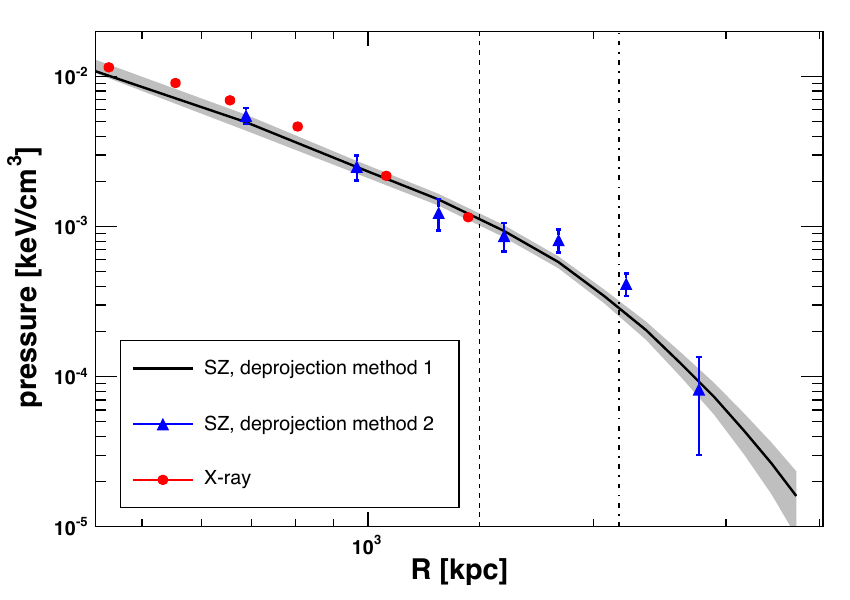}}
\caption{Electron pressure profile obtained using \emph{Planck} and \emph{XMM-Newton}. The blue data points and gray shaded area show the deprojected SZ electron pressure obtained using a non-parametric and a parametric method, respectively. The red data points denote the pressure profile obtained from X-ray spectral analysis.}
\label{fig:pressure}
\end{figure}

In Fig. \ref{fig:pressure} we show the \emph{Planck} pressure profile obtained using two different deprojection methods \citep[see][]{tchernin16}. The results are compared with the pressure profile measured from a spectral X-ray analysis. An excellent agreement between X-ray and SZ pressure profiles is found over the range of overlap. This confirms that X-ray and SZ techniques return a consistent picture of the gas properties in galaxy clusters and further validates the method that is put forward in X-COP.

\subsection{Joint X-ray/SZ analysis}

Once the gas density and pressure profiles are determined, the radial profiles of temperature $kT=P_{e}/n_{e}$ and entropy $K=P_{e}n_{e}^{-5/3}$ can be inferred. The gravitating mass profile can also be recovered by solving the hydrostatic equilibrium equation,

\begin{equation}\frac{dP}{dr}=-\rho\frac{GM(<r)}{r^2}.\label{eq:hydro}\end{equation}

By comparing the gravitating mass with the gas mass obtained by integrating the gas density profile, the profile of intracluster gas fraction can also be recovered. 

\begin{figure}[t]
\resizebox{\hsize}{!}{\includegraphics[width=0.5\textwidth]{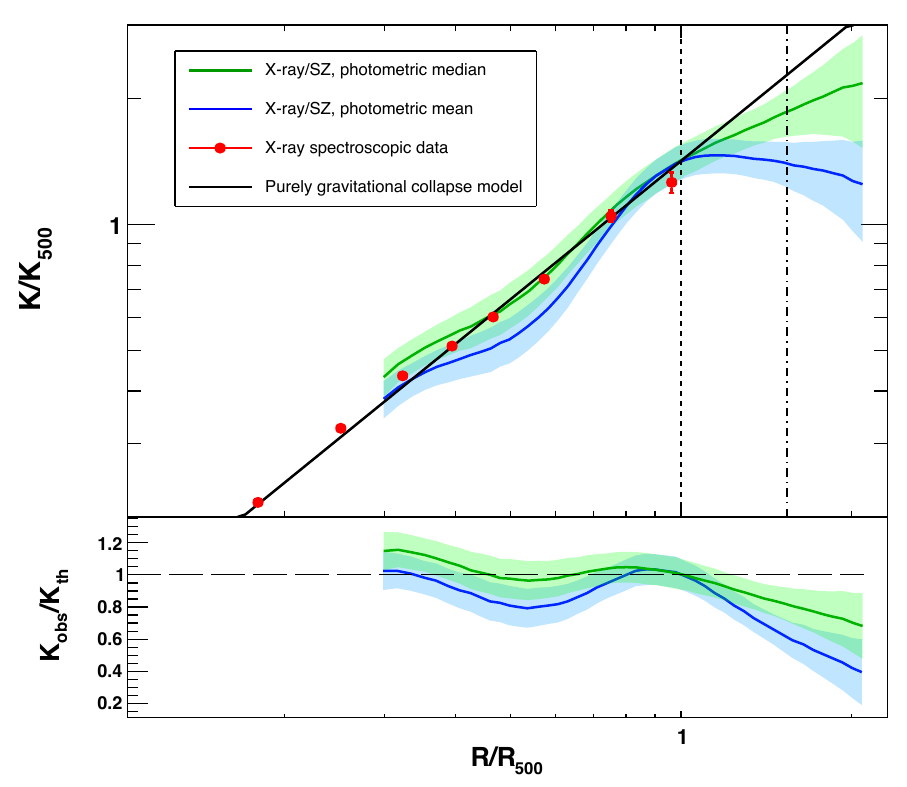}}
\caption{Entropy profiles of Abell 2142 obtained by combining \emph{Planck} and \emph{XMM-Newton} data. The green curve and shaded area show the entropy profile inferred from the median (clumping corrected) gas density profile, while for the blue curve the mean (biased) density profile was used. The red points show the spectroscopic X-ray measurements. The black curve represents the expectation of pure gravitational collapse \citep{voit05}.}
\label{fig:entropy}
\end{figure}

In Fig. \ref{fig:entropy} we show the radial entropy profile of Abell 2142 obtained by combining X-ray and SZ data. Once again, the data are compared with the results of the spectroscopic X-ray analysis, which highlights the much broader radial range accessible by the joint X-ray/SZ technique. The observed entropy profiles are compared with the prediction of numerical simulations using gravitational collapse only \citep{voit05}. 

Interestingly, when combining the SZ data with the median (clumping corrected) gas density profile, the recovered entropy profile is consistent with the theoretical expectation within $1\sigma$, whereas if the mean (biased) density profile is used, at large radii the entropy falls significantly below the expectations. In the latter case, the behavior of the entropy profile is very similar to a number of recent \emph{Suzaku} studies, which found a deficit of entropy beyond $R_{500}$. Our analysis thus highlights the importance of gas clumping when interpreting the results of \emph{Suzaku} observations of cluster outskirts.

\begin{figure}[t]
\resizebox{\hsize}{!}{\includegraphics[width=0.5\textwidth]{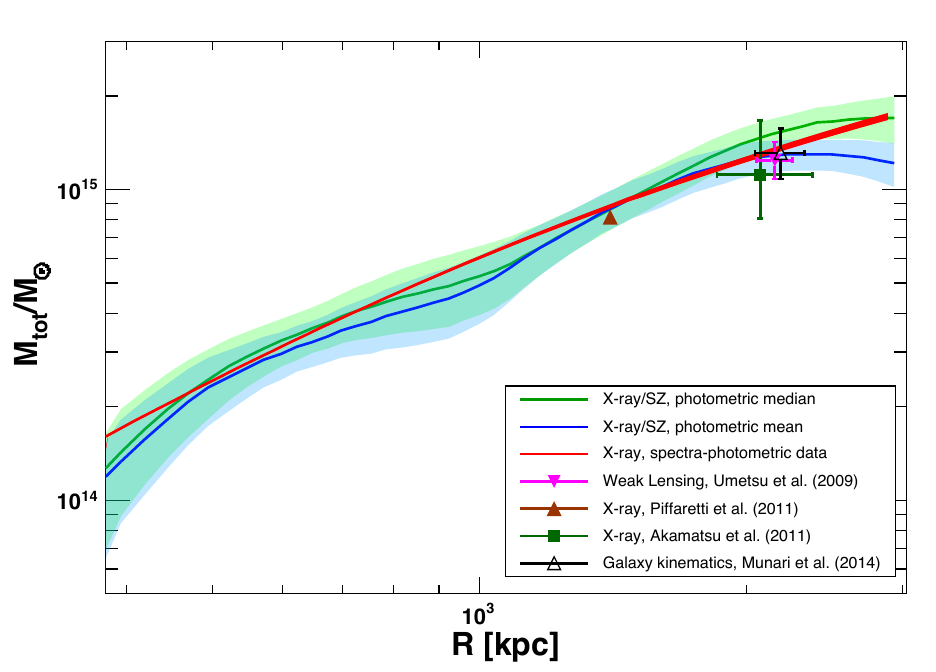}}
\caption{Mass profiles of Abell 2142 obtained by combining \emph{Planck} and \emph{XMM-Newton} data. The green curve and shaded area show the mass profile inferred from the median (clumping corrected) gas density profile, while for the blue curve the mean (biased) density profile was used. The red curve indicates the mass profile inferred from spectroscopic X-ray measurements \citep{ettori10}. The symbols indicate the values of $M_{200}$ calculated from weak gravitational lensing \citep[pink triangle,][]{umetsu09}, galaxy kinematics \citep[black triangle,][]{munari14} and \emph{Suzaku} X-ray \citep[green square,][]{akamatsu11}}
\label{fig:mass}
\end{figure}

In Fig. \ref{fig:mass} we show the gravitating mass profile obtained by solving the hydrostatic equilibrium equation (Eq. \ref{eq:hydro}). In this case, we find that the mass profiles calculated with the mean and median profiles are consistent. We measure $M_{200}=(1.41\pm0.03)\times10^{15}M_\odot$, in good agreement with the values calculated with weak gravitational lensing \citep[$1.24_{-0.16}^{+0.18}\times10^{15}M_\odot$,][]{umetsu09} and galaxy kinematics \citep[$1.31_{-0.23}^{+0.26}\times10^{15}M_\odot$,][]{munari14}. Thus, our data do not show any sign of hydrostatic bias even when extending our measurements out to $R_{200}$.

\section{Conclusion}

In this paper, we presented an overview of the \emph{XMM-Newton} cluster outskirts project (X-COP), a very large programme on \emph{XMM-Newton} that aims at performing a deep X-ray and SZ mapping for a sample of 13 massive, nearby galaxy clusters. The clusters of the X-COP sample were selected on the basis of their strong SZ effect in \emph{Planck} data. The combination of X-ray and SZ data over the entire volume of X-COP clusters will allow us to improve our knowledge of the intracluster gas out to $R_{200}$ and beyond, in order to reach the following goals: \emph{i)} measure the radial distribution of the thermodynamic properties of the ICM; \emph{ii)} estimate the global non-thermal energy budget in galaxy clusters; \emph{iii)} detect infalling gas clumps to study the virialization of infalling halos within the potential well of the main structure. 

We presented a pilot study on the galaxy cluster Abell 2142 \citep{tchernin16}, demonstrating the full potential of X-COP for the study of cluster outskirts. The cluster is detected out to $2\times R_{500}\sim R_{100}$ both in X-ray and SZ. The two techniques provide a remarkably consistent picture of the gas properties, and they can be combined to recover the thermodynamic properties of the gas and the gravitating mass profile out to the cluster's boundary. Our results highlight the importance of taking the effect of gas clumping into account when measuring the properties of the gas at large radii, where accretion from smaller structures is important. In the near future, X-COP will bring results of similar quality for a sizable sample of a dozen clusters, allowing us to determine universal profiles of the thermodynamic quantities and gas fraction out to the virial radius.

\acknowledgements
Based on observations obtained with \emph{XMM-Newton}, as ESA science mission with instruments and contributions directly funded by ESA Member States and NASA. The development of \emph{Planck} has been supported by: ESA; CNES  and  CNRS/INSU-IN2P3-INP  (France);  ASI,  CNR,  and  INAF  (Italy); NASA  and  DoE  (USA);  STFC  and  UKSA  (UK);  CSIC,  MICINN,  JA  and RES (Spain); Tekes, AoF and CSC (Finland); DLR and MPG (Germany); CSA (Canada);  DTU  Space  (Denmark);  SER/SSO  (Switzerland);  RCN (Norway); SFI (Ireland); FCT/MCTES (Portugal); and PRACE (EU). 

%\newpage%%%%%%%%%%%%%%%%%%%%%%%%%%%%%%%%%%%%%%%%%%%%%%%%%%%%%%

\bibliographystyle{aa}
\bibliography{eckert_xmm}

\end{document}